\documentclass[letterpaper]{appolb}
\usepackage{epsfig}
\usepackage[colorlinks]{hyperref}

\catcode`@=11 
\newdimen\z@ \z@=0pt 
\newskip\z@skip \z@skip=0pt plus0pt minus0pt
\def\m@th{\mathsurround=\z@}
\def\ialign{\everycr{}\tabskip\z@skip\halign} 
\def\eqalign#1{\null\,\vcenter{\openup\jot\m@th
  \ialign{\strut\hfil$\displaystyle{##}$&$\displaystyle{{}##}$\hfil
      \crcr#1\crcr}}\,}
\catcode`@=12 

\newcount\equanumber         \equanumber=0
\def\eqname#1{\relax \global\advance\equanumber by 1%
          \xdef#1{{\number\equanumber}}}

\def\eqn#1{\eqno\eqname{#1}(#1)\hypertarget{#1}{}}
\def\hyrefq#1{\hyperlink{#1}{(#1)}}
\def\hyref#1{\hyperlink{#1}{#1}}
\newcount\figurecount     \figurecount=0
\def\FIG#1{ \global\advance\figurecount by 1 \xdef#1{\the\figurecount}}
\newcount\tablecount     \tablecount=0
\def\TAB#1{ \global\advance\tablecount by 1 \xdef#1{\the\tablecount}}
\let\cl=\centerline
\def\allcap#1;#2;{{\renewcommand{\baselinestretch}{.9}\captionfont
\newdimen\fcwidth \fcwidth=\textwidth \advance\fcwidth by -2cm
\setbox0=\hbox{{\bf Fig. #1.} #2}  
  \ifdim \wd0>\fcwidth  
       \vbox{\noindent
          \parshape=1 1truecm \fcwidth {\bf Fig. #1.} #2}
    \else
       \cl{{\bf Fig. #1.} #2}
    \fi} }

\def\figbox#1;#2;{\parbox{#2cm}{%
\vglue3mm\epsfig{file=\figdir#1.eps,width=#2cm}\vglue3mm}}
\def\figboxc#1;#2;{\cl{\figbox #1;#2;}}

\def\ie{{\it\kern-2pt i.\kern-.5pt e.\kern-2pt}}  
\def\BR{\hbox{BR}}  
\def\up#1{$^{#1}$}  \def\dn#1{$_{#1}$}
\def\ifm#1{\relax\ifmmode#1\else$#1$\fi}
\def\Bbar{\ifm{\rlap{\kern.22em\raise1.9ex\hbox to.58em{\hrulefill}} B}}
   \def\deg{\ifm{^\circ}}
 \def\bra#1;{\ifm{\langle\,#1\,|}} 

\def\to{\ifm{\rightarrow}} \def\sig{\ifm{\sigma}}   \def\plm{\ifm{\pm}}
  
\def\ff{$\phi$--factory}  \def\DAF{DA\char8NE}  
\def\f{\ifm{\phi}}  \def\klo{KLOE}
\def\pb{{\bf p}} \def\pic{\ifm{\pi^+\pi^-}} \def\pio{\ifm{\pi^0\pi^0}} 
\def\pim{\ifm{\pi^-}}  \def\pip{\ifm{\pi^+}} \def\po{\ifm{\pi^0}}

   \def\epem{\ifm{e^+e^-}}
\def\km{\ifm{K^-}}  \def\kp{\ifm{K^+}} \def\po{\ifm{\pi^0}}
\def\ks{\ifm{K_S}} \def\kl{\ifm{K_L}} 
  
\def\eps{\ifm{\epsilon}} 
    
\def\Kb{\ifm{\rlap{\kern.2em\raise1.9ex\hbox to.6em{\hrulefill}} K}}
\def\kpm{\ifm{K^\pm}}  
\def\C{\ifm{C}}  \def\P{\ifm{P}}   \def\CP{\ifm{CP}} 
\def\noc{\relax\hglue0pt{\rlap{$C$}\raise.15ex\hbox{$\kern
.18em\backslash$}}}
\def\nop{\relax\hglue0pt{\rlap{$P$}\raise.15ex\hbox{$\kern
.18em\backslash$}}}
\def\noT{\relax\hglue0pt{\rlap{$T$}\raise.15ex\hbox{$\kern
.18em\backslash$}}}
  
\def\ko{\ifm{K^0}}  \def\kob{\ifm{\Kb\vphantom{K}^0}}
\def\gam{\ifm{\gamma}} \def\kkb{\ifm{\ko\kob}} 
 \def\ab{\ifm{\sim}}  \def\x{\ifm{\times}}
\def\sta#1;{\ifm{|\,#1\,\rangle}} \def\ket#1;{\ifm{|\,#1\,\rangle}} 
\def\L{\ifm{{\mathcal L}}}  \def\R{\ifm{{\cal R}}}
\def\pt#1,#2,{\ifm{#1\x10^{#2}}}

\def\ord#1;{\ifm{{\mathcal O}(#1)}}
\def\bye{\end{document}}

\begin{document}

\def\figdir{}
\pagestyle{plain}
\newcount\eLiNe\eLiNe=\inputlineno\advance\eLiNe by -1
\title{A Flavor of KLOE\\
\\
}
\author{Juliet Lee-Franzini
\address{Laboratorio Nazionali Di Frascati Dell'INFN
Frascati, Rome, Italy.}
\and
Paolo Franzini
\address{Sezione di Roma1 dell'Istituto Nazionale di 
Fisica Nucleare, Roma, Italy.}}
\maketitle

\begin{abstract}
This paper is a simple, quick guide to KLOE, the flagship experiment of INFN's $\phi$ factory \DAF\ at Frascati. \klo 's design principles, properties, its physics accomplishments and its impact on ``flavor physics'', are described in terms comprehensible to non specialists.
\end{abstract}
\def\captionfont{\small}

\FIG\dafne
\FIG\adone
\FIG\kloe
\FIG\kloehall

\section{Introduction}
In late 1988 the Italian Institute for Nuclear Physics, INFN, decided to construct an \epem\ collider meant to operate around 1020 MeV, the mass of the \f-meson. The \f-meson decays mostly to kaons, neutral and charged, in pairs. The \f\ production cross section peaks at about 3 $\mu$barns. \f-mesons decay \ab49\% of the time into \kp\km pairs and 34\% of the time into \kkb pairs. Even with a modest luminosity, \L=100 $\mu$b\up{-1}/s, hundreds of kaon pairs are produced per second.  This collider, called a \ff\ and christened \DAF, is located at the Laboratori Nazionali di Frascati, LNF, INFN's high energy physics laboratory near Rome \cite{DAFNE+90:prop}. 
\subsection{\DAF}
The \DAF\ layout is shown in fig. \hyref\dafne. 
The heart of the collider are two storage ``rings'' in which 120 bunches of electrons and positrons are stored. 
Each bunch collides with its counterpart once per turn, minimizing the perturbation of each beam on the other. 
Electrons and positrons are injected in the rings at final energy, \ab 510 MeV. Electrons are accelerated to final energy in the Linac, accumulated and cooled in the accumulator and transferred to a single bunch in the ring. Positrons require first accelerating electrons to about 250 MeV to an intermediate station in the Linac, where positrons are created. The positrons then follow the same processing as electrons.

\vbox{\hypertarget{\dafne}\break\cl{\figboxc fig01;11.0;}}
\allcap\dafne;The \DAF\ complex.;\vglue2mm\noindent
Since the stored beam intensities decay rapidly in time, the cycle is repeated several times per hour. A collider is characterized by its luminosity \L\ defined by: event rate=\L\x cross section, beam lifetime and beam induced radiation background.

\vbox{\hypertarget{\adone}\break\cl{\figboxc adone;11.8;}
\allcap\adone;The Adone building.;}\vglue2mm\noindent
The latter is large for short lifetimes, which also result in the average luminosity being smaller than the peak value, in a word, lower event yield and large, variable background. \DAF\ is housed in the ADONE building at LNF, see fig. \hyref\adone.

\subsection{KLOE}
INFN soon realized that a state-of-the-art detector was needed to collect data at \DAF, from which physics results would materialize. Thus a couple of years later, in summer 1991, the KLOE Collaboration was initiated, though it was not formally approved nor fully funded for another couple of years, because the authorities had not envisioned fully the magnitude of \klo. By late 1998 however, \klo\ was designed, constructed, completely tested, and the humongous KLOE detector, complete with all of its electronics for signal processing, event gathering and transmission, was moved from its own specially built assembly hall onto the newly commissioned \DAF\ 's South Interaction Region. In the subsequent seven years KLOE, after beginning with a trickle of beam, by 2006, had produced data of such precision that the kaon, the first fundamental particle to introduce the concept of ``flavor" into our current way of thinking, got its definitive 21st century portrait re-mapped with high precision, 60 years after its discovery.

The KLOE saga, with its concomitant joys and travails, would be an entertaining tale that should be written up somewhere. Our purpose in this paper, instead, is to give the non-specialist reader a``flavor'' of the KLOE experiment, in order to better appreciate the role that a complete experiment such as KLOE plays in pushing forward the physics frontier.
\section{DETECTOR}
The KLOE detector has a giant inner core, approximately 125 meters cubed, see fig. \hyref\kloe , and was built in Frascati in the KLOE assembly building, see fig. \hyref\kloehall .
It is on the same scale and complexity as the ``general purpose detectors'' of its time that were operating at the world's foremost laboratories such as LEP at CERN. If we consider that LEP operated at about one hundred times \DAF's energy, producing much more energetic particles and with correspondingly greater multiplicities, \klo's large size would seem unwarranted. We shall see however that while KLOE's complexity was necessitated  by its intent of being a definitive high precision experiment, its size, instead, was dictated by the quirk of one of the two neutral kaons, called the \kl, of living an extraordinarily long time (for short lived particles) , about 51 nanoseconds. Kaons from \f-mesons decaying at rest travel at approximately one fifth of the speed of light. This low speed is a great bonus for many reasons. However the mean path travelled by a \kl-meson, $L_{\kl}=\gam\beta c\tau$ is 3.48 m. Thus if we want a detector which can catch $1-1/e$=63\% of all decaying neutral long-lived kaons, it must have a radius of some three and a half meters. We compromise on two meters and catch some 40\% of those decays. Less than that in fact, since some volume is inevitably lost around the the beam interaction point, IP, and some more is lost at the outer edge, in order to be able to recognize the nature of the decays close to it.

\vbox{\hypertarget{\kloe}\break\cl{\figboxc fig03;9.9;}
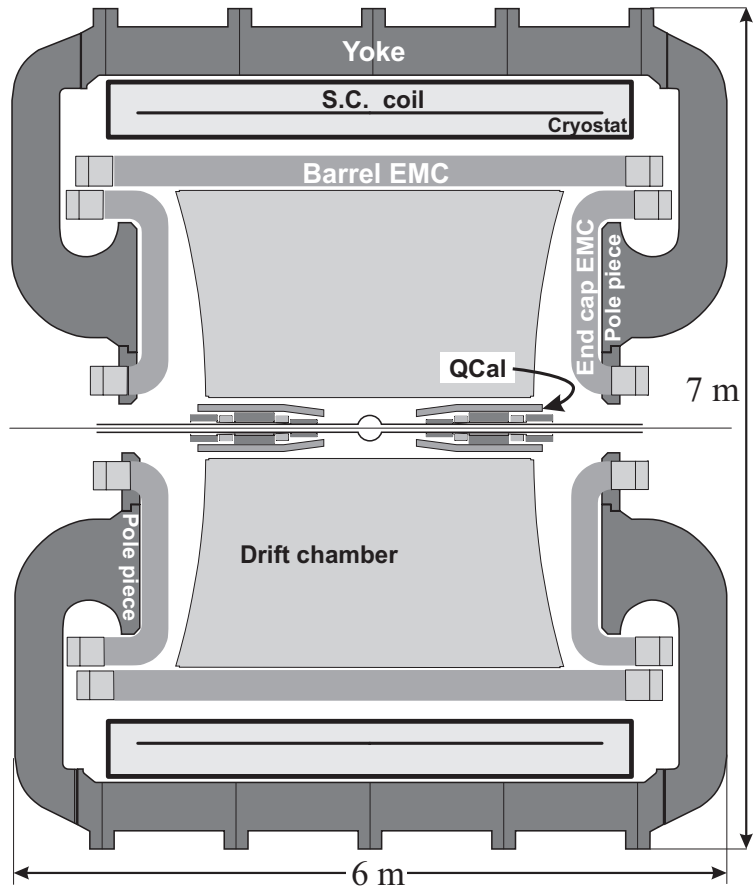
\allcap\kloe;Cross-sectional view of the KLOE experiment, showing the interaction region, the drift chamber (DC), the electromagnetic calorimeter (EMC), the superconducting coil, and the return yoke of the magnet.;}

\subsection{Drift Chamber}
Catching a decay really means observing the decay products or particles. 
Charged particle travelling in a medium, a gas, undergo repeated collisions with electrons. As a result a string of ion pairs marks its passage. This is how KLOE came to build the largest drift chamber, DC, in high-energy physics. A DC is a gas filled chamber traversed by a large number of wires, some kept at +2000 volt, anode wires, some at ground. Electrons of the ion pairs created along the particle trajectory drift to the positive voltage wires. Thanks to an avalanche multiplication mechanism a detectable signal appears at the wire's end. Measuring the drift time locates the distance of the track from the anode wires and ultimately we can reconstruct the trajectory of the particles. Sophisticated  electronics mounted at the the wire's end senses the tiny signals. The \klo\ DC has \ab52,000 wires, about 12,500 of them are sense wires, the remaining ones shape the electric field in 12,500 small (2x2, 3x3 cm\up2) cells. For a particle crossing the entire chamber we get dozens of space points from which we reconstruct a track. The DC is surrounded by a superconducting coil that produces a magnetic field of \ab0.5 Tesla. The particle trajectories curl up and we determine the particle's momentum from the curvature of its track with a fractional accuracy of \ab0.4\%. The spatial resolution is well below 200 $\mu$m in the ``$\phi$-coordinate'' \cite{DCNIM}.

\vbox{\hypertarget{\kloehall}\break\cl{\figboxc kloehall;11;}
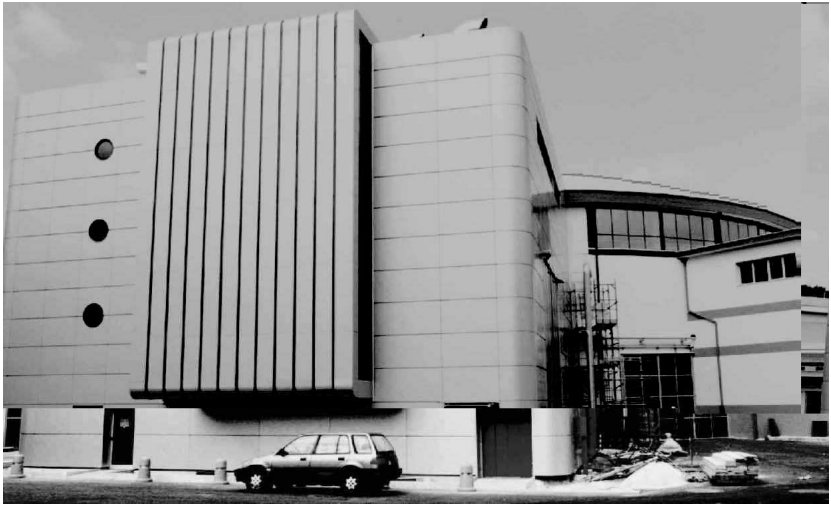
\allcap\kloehall;KLOE Assembly Hall. The KLOE control room is on the top floor. The Adone cupola is just visible behind.;}

\subsection{Electromagnetic Calorimeter}
The neutral decay products of the \kl's, its short-lived partner \ks\  (lifetime \ab\ 89 picoseconds) and of those of their charged brethren, the \kpm's, which we need to detect, are photons, either directly produced or from the decay of neutral pions into two photons. Photons do not ionize, but interact with matter by photoelectric effect, Compton scattering and for energies larger than a few MeV (as in most of our cases) by electron positron pair production. Electrons and positrons in turn radiate photons when traversing matter, a process quite similar to pair-production. Thus, one readily imagines that a photon, or an electron for that matter, travelling through dense, high $Z$  matter, repeatedly undergoes radiation and pair production in cascade, until all its energy is expended in a so called shower of \epem\ and photons. Detection of these particles and their energy measures the original photon's (or electron's) energy. This is done with a ``calorimeter'', which also usually pin points the position of the electromagnetic cascade (or shower) as well. 

The \klo\ calorimeter EMC, is a sampling calorimeter. It is made of lead passive layers that accelerate the showering process and of scintillating fiber sensing layers. Energy densities are sampled at different depth along the shower development. The basic structure of the \klo\ calorimeter is a composite of very thin, 1 mm, groved Pb layers between which are embedded 1 mm diameter scintillating fibers. \ab 200 Pb-fiber layers are used. The high ratio of active material to radiator and the very fine subdivision result in better energy resolution. In addition, the special care in design and assembly of the Pb-fiber composite ensures that the light propagates along the fiber in a single mode, resulting in a greatly reduced spread of the light arrival time at the fiber ends. Photomultiplier tubes at both fiber ends transform this light into electric pulses. Their amplitude gives the amount of energy deposited. The arrival times locates the position of the shower to ${\cal O}$(1) cm accuracy. The resulting material has a radiation length $X_0$ of 1.5 cm. 30 cm or 15 $X_0$ fully contain a shower of 500 MeV. About 5000 photomultiplier tubes view the 24+32+32 modules into which the calorimeter is subdivided. 

The calorimeter material is formed into a barrel, aligned with the beams and surrounding the DC. In additions two end plugs, draped over the magnet pole pieces hermetically close the calorimeter \ab98\% of 4$\pi$, see fig. \hyref\kloe. The KLOE calorimeter, EMC, with time resolution of the order of 54 picoseconds for a one GeV photon, is jocularly referred to as ``the fastest calorimeter in the West". The energy resolution of \klo\ is also excellent, of the order of 5.7\% for a one GeV photon \cite{EmCNIM}.

\FIG\beamp
\subsection{Beam Pipe}
At \DAF\ the electron and positron beams collide almost head on, $\theta=\pi-25$ mrad. \f-mesons are therefore produced with \ab 13 MeV/c, in the horizontal plane, toward \DAF's center. The \ks's with its lifetime of \ab 89 ps or 6mm path length, emit secondary particles almost isotropically from the interaction point (IP). At the IP the beam pipe is built of an Al-Be alloy 1 mm thick and shaped as a 10 cm radius sphere, fig. \hyref\beamp.\\
\vbox{\hypertarget{\beamp}\break\cl{\figboxc beamp3;8.8;}
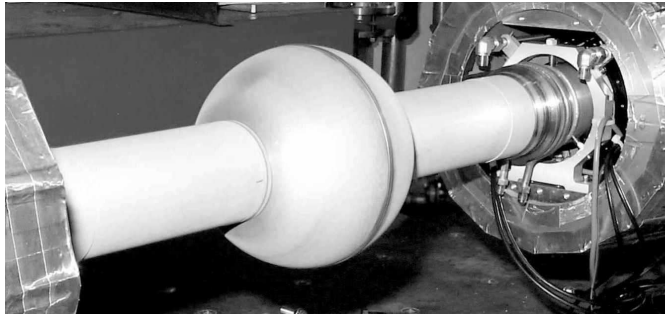
\allcap\beamp;Beam pipe at the KLOE interaction region. Just visible on either side are the final focus quad ends.;}\vglue1mm\noindent
The \ks\ amplitude is reduced by a factor of \ab5,000 at the sphere surface. Regeneration, \kl\to\ks, can be neglected in the crucial interference region.

\subsection{Trigger}
The total \epem\ cross section at 1020 MeV is \ord100; $\mu$b, of which only 3 $\mu$b is due to \f-production. 
In \DAF\ the collision frequency reaches up to \pt3.68,8, Hz. A trigger is necessary to fire up the electronics, to process signals from some 20,000 detector elements, only when something interesting happened. We want $\phi$ decays, and a fraction \epem\ \to\ \epem\ (Bhabha) or \gam\gam\  events, for \L\ measurement and for continuous monitoring and calibration of \klo. We also wish to ignore cosmic rays descending into \klo\ from the heavens. An elaborate electronic logic system rapidly recognizes topologies and energy releases of interest, providing the signal or ``trigger'' that something interesting happened. The KLOE trigger is based on isolated energy releases in the calorimeter and the presence of  \ab 15 hits in the DC within a time window of 250\ab ns from beam crossing. This trigger, called Level 1, initiate conversion in the front-end electronics modules, which are subsequently read out following a fixed time interval. A second level trigger, requiring \ab120 hits within a 1.2-$\mu$s window in the DC validates the trigger. The \klo\ trigger also recognize and ignores cosmic ray events \cite{trigNIM}. A valid trigger initiates read out. The trigger system also provides a fast measurement of the \DAF\ luminosity. A precision measurement of the luminosity is obtained off-line from Bhabha events \cite{lumi}.

\FIG\kkevent
\subsection{Data Acquisition}

At a luminosity of \ab100 $\mu$b\up{-1}/s, events worth keeping occur at \ab2200 Hz. Of these \ab300~Hz are from \f\ decays. All digitized signals from the calorimeter, drift chamber, calibration and monitoring systems are fed from the electronics that sit on platforms mounted on the \klo\ iron yoke, via fiber optics to the \klo\ computers in the next building. In the near-by \klo\ control room one can see real time displays of  reconstructed charged tracks traversing the DC and photons depositing energy clusters in the calorimeter. The display of two events is shown in fig, \hyref\kkevent.

The events are reconstructed, meaning that charged tracks and energy deposits that occurred at the same instant in time on the various subcomponents of \klo\, using constantly updated calibration files that monitor both the time and energy scales, are associated together, properly labelled and packed into a unique file. Furthermore, background hits that accidentally occurred during the time an event time window is open, are also recorded such that these hits can be added on (superimposed) on the simulated \klo\ events (MC). The \klo\ MC simulation program reproduces the \klo\ geometry, physical responses and trigger requirements extremely well, but of course cannot foretell the moment to moment variation of the background.

All raw data, background, reconstructed and MC events are recorded on tape. \klo\ has a huge tape library of \ab800~TB capacity \cite{DAQNIM}. At reconstruction time, events are classified into broad categories or ``streams''. Data summary tapes are produced as a separate procedure.
\vglue1mm
\vbox{\hypertarget{\kkevent}\break\cl{\figbox klksnocp2; 5.7;\kern.3cm\figbox klkspio-bw;5.7;}
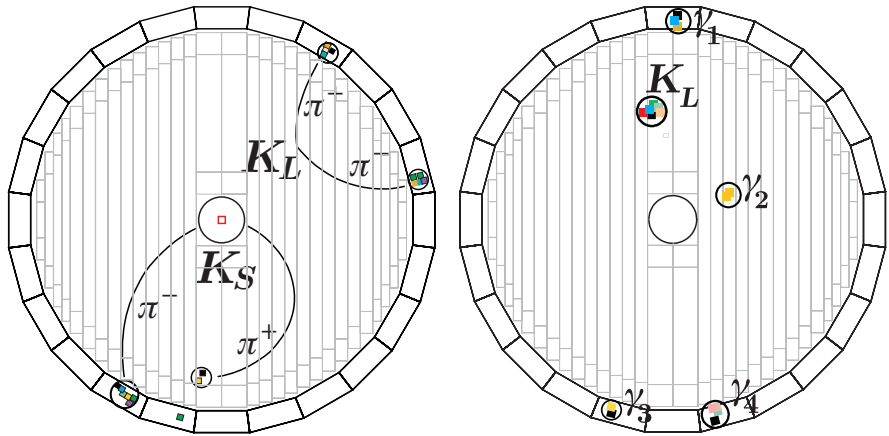
\allcap  \kkevent; \ks\kl\ events in \klo. Left: both \kl\ and \ks\ decay to charged pions \pip\pim. Color patches indicate energy in the calorimeter. Right: 5 neutral particles. \gam\dn1 trough \gam\dn4 are showers from 4 photons from \ks\to\pio\to4\gam. The \kl-mesons reaches the calorimeter before decaying. It literally crashes in the EMC, producing a huge energy release.;}\vglue1mm
\noindent
Any of the \klo\ analyses, even after event reconstruction is quite complex. It includes making data summary ``tapes'' and extensive modelling of the detector response. This requires generation of large number of Monte Carlo (MC) events for each studied process. Both data and MC summaries (\ab80~TB) are cached on disk for analysis \cite{offlineNIM}.

\FIG\klp
\subsection{\kl\ Path}
The production of \ks\kl\ pairs in \f-decays together with the excellent timing accuracy of the \klo\ EMC allows us to locate \kl\to\pio\ decay point, in the terrible jargon of today's physicists, a ``neutral vertex'', to accuracies of \ord3; mm, as illustrated in figure \hyref\klp.
This procedure is of fundamental importance in most of \klo\ measurements.

\vbox{\hypertarget{\klp}\break\cl{\figboxc kl-path;6;}
\allcap\klp; $D$ is the \kl\to\pio\to4\gam\ decay point. The \kl\ path is obtained from the time for 
each photon to arrive at the calorimeter, $t$(I \to A), and the \ks\ direction.;}\vglue1mm

\FIG\bkl

\subsection{The \ks\ beam}
\label{ksbeam}
The two neutral kaon state from a \f-decay is given, in the \f\ rest frame, by
$$\eqalign{\sta i;&={\sta\ko,\pb;\sta \kob,-\pb;-\sta\kob,\pb;\sta\ko,-\pb;\over\sqrt2}=\cr
&\kern3cm{\sta\ks,\pb;\sta \kl,-\pb;-\sta\kl,\pb;\sta\ks,-\pb;\over\sqrt2}.\cr}\eqn\twok$$
Therefore observation of a \kl(\ks) decay ensures the presence of a \ks(\kl) meson, travelling in the opposite direction. This provides the unique opportunity of preparing a pure \ks-beam. A pure \kl\ beam can thus also be defined, although most experiments accomplish the same by simply waiting for the \ks-amplitude to die away. We refer to the process of defining a \ks\ or \kl\ sample as tagging: observation of a \kl(\ks) decay tags the presence of a \ks(\kl) meson of known momentum {\bf p}.
Because of the exceptional timing capabilities of the \klo\ EMC and the slowness of the kaons from \f-decays, \klo\ tags \ks-mesons in a unique way.
As already mentioned, kaon have a velocity $\beta=0.2162$ in the \f\ frame and $\beta$=0.193 to 0.239 in the lab and 60\% of them reach the calorimeter in a time $\geq$31 ns. \kl-mesons interact in the calorimeter with an energy release up to their mass, 497 MeV. For an energy release of 100 MeV \klo\ has a time resolution of \ab0.3 ns, \ie\ a velocity accuracy of \ab1\%. The accuracy in measuring the \kl\ velocity, before exact determination of the collision time, is shown in fig. \hyref\bkl.\\
\vbox{\hypertarget{\bkl}\break\cl{\figboxc betakl0;7.8;}
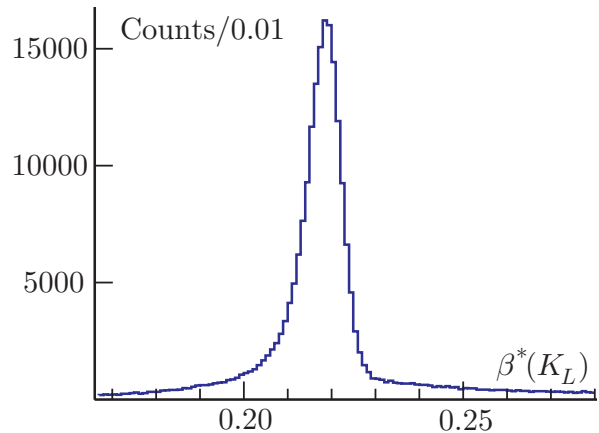
\allcap\bkl; \kl\ velocity in \f-frame from time of flight (before exact  determination of the collision time).;}\vglue1mm
\noindent
In \klo, we call a \kl\ interaction in the EMC a ``k-crash''.
Note that the above error on $\beta$(\kl) corresponds to an error on the kaon energy of \ab0.25 MeV or 0.5 MeV for the collision energy, with just one event. We use this method, together with \epem\to\epem\ events to monitor the \DAF\ energy and energy spread.

\section{Kaon Physical Parameters and Decays}
Since the kaon's discovery about sixty years ago, knowledge of its properties, from masses, lifetimes, and how often each species decays into to which particles, has been accumulated piecemeal literally over hundreds of individual experiments, performed at various laboratories all over the world. Each experiment has its own peculiarity, contingent upon the particle beam available and its own equipment's acceptance range, which can result in highly accurate measurements from a statistical point of view (namely based on thousands of data events) but with hidden corrections peculiar to it perhaps unknown even to its authors. The particle physics ``self-nominated authority'' in charge of compiling the results, often adopt a ``democratic'' way of weighing the various results according to their nominal errors and literally applying a fudge factor when faced with two inconsistent measurements. While this procedure skirts the problems of judging the relative accuracy of the results obtained over time, it sometimes ends up with a hodge-podge of inconsistent results forced into a mold of dubious scientific validity.

The mission of \klo\ is to measure most, if not all, of the properties of the kaon system to high accuracy, with a single detector simultaneously.
Moreover, radiation is always present and calculating its effects involves cancellation of infinities. Only if experimental answers are fully inclusive with respect to radiation, can they be compared with calculation in a model. Most branching ratios reported in the past have been very vague on this point. \klo\ has been particularly attentive to the inclusion of radiation.

\klo\ can accomplish the said goal because as described before, 1, it has available tagged, pure kaon beams (the only \ks\ one) beams of precisely known momenta ( known event by event, from kinematic closure) and 2, of known flux, the number created of each specie, 3, and catches all the decay products hence determine the fractions of decays into different channels. In short, \klo\ can measure each kaon specie's total decay rate $\Gamma$, the inverse of which is the lifetime, and its fractions of decay into different channels, called branching ratio, BR.

\subsection{Masses}
Kaons masses are about 500 MeV, very close to one half the \f-meson mass, known to an accuracy of 0.019/1019.460 \ab\pt2,-5, by the use of the $g$-2 depolarizing resonances, \cite{pdg06}. In \klo\ we use the \ks\ decay into \pio\ and \pic\ routinely to monitor the EMC and DC performance. Measuring the \ks\ mass requires measuring $M(\f)-2 M(\ko)$=12.05 MeV, corresponding to $p(\ko)$=110.2 MeV.  Given the billions of \ks\to\pic\ decays available in KLOE, the statistical error can be arbitrarily small. The accuracy is ultimately limited by uncertainties in the calculations of radiative corrections. 
\klo\ measures $M(\kl)=497.583\pm0.021$~  MeV \cite{llwidth}.

\subsection{\ko\ lifetimes}
\label{sec:KLlife}

Precision measurements of the lifetime of neutral kaons, especially those with many multibody decay modes is particularly difficult since it is in general not possible to prepare monochromatic beams of neutral particles nor to stop them. \klo\ enjoys the availability of such monochromatic beams. While the \ks\ lifetime, for many obvious reasons, is quite well known, \pt(0.8958 \pm0.0005),-10, s, this is not the case for the \kl\ lifetime.
\klo\ can measure the lifetime by observing the time dependence of the decay frequency to a single mode. Because of tagging, the lifetime can also be obtained by simply counting the total number of decays in a given time interval from a known \kl sample.

We have collected a sample of \ab 14 million \kl\to\pic\po\  decays over a 6-26 ns interval (\ab 40\% of the \kl\ lifetime). A fit of the proper time distribution gives $\tau_{\kl}=50.92\pm0.30$~ns \cite{kllife}.
From the second method, using \ab2 million \kl decays to all major channels, KLOE finds $\tau_{\kl}=50.72\plm0.36$~ns. The two determinations are in agreement and are almost entirely uncorrelated \cite{klbr}. The combined result is 50.84\plm0.23 ns, compared to 51.54\plm0.44~ns from 1972.

\subsection{\kpm\ lifetime}

The measurements of the \kpm\ lifetime listed in the PDG compilation \cite{pdg06} exhibit poor consistency. The PDG fit has a confidence level of \pt1.5,-3,, and the error on the recommended value is enlarged by a scale factor of 2.1. KLOE can measure the decay time for charged kaons in two ways. The first method is to obtain the proper time from the kaon path length in the DC. Oue preliminary value for the lifetime is $\tau_{\kpm}$ = $12.367 \pm 0.044 \pm 0.065$ ns, where the first error is statistical and the second systematic.

The second method is based on the precise measurement of the arrival times of the photons from $\kpm\to\pi^\pm\pi^0$ decays and the determination of  the decay point from the track vertex in the DC. It is near completion. Both methods provide results accurate to the level of 0.1\%.

\subsection{\ks\ decays}

As already mentioned, central to \klo\ 's \ks\ studies is the use of \kl-crash tag: the position of the \kl\ crash, together with the kinematics of the $\f\to\ks\kl$ decay, determines the trajectory of the \ks\ with a momentum resolution of approximately 1 MeV and an angular resolution of better than 1\deg. Using \kl-crash tag \klo\ was able to measure the \ks\ branching ratios whose values span six orders of magnitude.\vglue4mm

\noindent{\mathversion{bold}$\ks\!\to\!\pic(\gamma),\ \pio\kern8mm-$} \kern8mm
The \ks\ decays overwhelmingly (99\%) into two pions: \pio and \pic. The ratio of the charged decay mode to that of the neutral $\R_{\pi}\equiv\Gamma(\ks\to\pic(\gam))/\Gamma(\ks\to\pio)$ is a fundamental parameter of the \ks\ meson. It is used in the extraction of values for phenomenological parameters of the kaon system, such as the differences in magnitude and phase of the $I=0,\,2$ $\pi\pi$ scattering amplitudes. It and the corresponding ratio from the \kl\ together determine the amount of direct CP violation in $K\to\pi\pi$ transitions. During the last six years \klo\ has done two precise measurements of $\R_{\pi}$, the final combined value for  $\R_{\pi}$ is $2.2549 \pm 0.0054$, a two part per mil measurement \cite{ks2pi}.\vglue3mm

\FIG\kslep     \TAB\ksdata 

\noindent{\mathversion{bold}$K_S\rightarrow \pi e\nu(\gamma)\kern8mm-$} \kern8mm
The \ks\ decays semileptonically, (meaning that leptons are present amongst its decay products) less than one per cent of the time. To pick out such decays where the event contains an unseen neutrino is non-trivial. Yet, KLOE has isolated a very pure sample of \ab13,000 semileptonic \ks\ decays and accurately measured the BRs for $\ks\to\pi^+e^-\bar\nu(\gam)$ and $\ks\to\pi^-e^+\nu(\gam)$ \cite{kslept}.

The basic steps in the analysis are: tag \ks\ decays by the \kl\ crash, make a cut on the $\pi\pi$ invariant mass (this removes 95\% of the $\ks\to\pic$ decays and reduces the background-to-signal ratio to \ab80:1), impose several geometrical cuts to further improve the purity of the sample and, in particular, remove contamination by events with early $\pi\to\mu\nu$ decays. Finally, stringent requirements are imposed on the particle's time of flight (TOF), which very effectively separates electrons from pions and muons and allows charge assignment of the final state.

Fig. \hyref\kslep\ shows the signal peak and the residual background in the
distribution of $\Delta_{Ep} = E_{\rm miss}-|{\bf p}_{\rm miss}|$
for the $\pi^-e^+\nu$ channel, where $E_{\rm miss}$ and ${\bf p}_{\rm miss}$
are respectively the missing energy and momentum at the vertex, evaluated 
in the signal hypothesis. For signal events, the missing particle is a
neutrino and $\Delta_{Ep}=0$.\\
\vbox{\hypertarget{\kslep}\break\cl{\figboxc  fig07;7.9;}
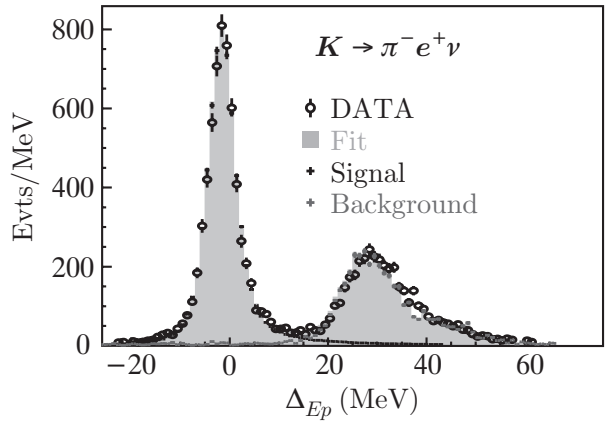
\allcap\kslep;{$\Delta_{Ep}$ distribution for $\ks\to\pi^-e^+\nu$ candidates, showing signal and background.};}\vglue1mm
\noindent
The numbers of $\pi e\nu$ decays for each charge state are normalized to 
the number of \pic\ events observed, resulting in the ratios in the first 
column of table \ksdata.
These ratios give the first measurement of the 
semileptonic charge asymmetry for the \ks:
$$A_S = \pt(1.5\pm9.6\pm2.9),-3,.$$
Using the result for $\R_\pi$\ of the previous section (also in table \ksdata), 
the absolute BRs for $\ks\to \pi\pi$ and $\ks\to\pi e\nu$
in the second column of the table are obtained.\vglue2mm
\def\vst{\vphantom{\vrule height4mm depth0pt}}
\def\vsta{\vphantom{\vrule height5mm depth0pt}}
\begin{center}
\cl{Table \ksdata. KLOE measurements of $K_S$ branching ratios}\vglue2mm
\begin{tabular}{@{}lcc@{}}
\hline\hline
\vst Mode & BR(mode)/BR(\pic) & BR(mode) \\
\hline
\vsta $\pic$ & --- & ($69.196\pm0.024\pm0.045$)\% \\
\vst $\pio$ & 1/($2.2549\pm0.0054$) & ($30.687\pm0.024\pm0.045$)\% \\
\vst $\pi^-e^+\nu$ & \pt(5.099\pm0.082\pm0.039),-4, & \pt(3.528\pm0.057\pm0.027),-4,\\
\vst $\pi^+e^-\bar{\nu}$ & \pt(5.083\pm0.073\pm0.042),-4, & \pt(3.517\pm0.050\pm0.029),-4, \\
\vst $\pi e \nu$ & \pt(10.19\pm0.11\pm0.07),-4, & \pt(7.046\pm0.076\pm0.051),-4,\\
\hline
\end{tabular}
\end{center}\vglue2mm\noindent

\noindent{\mathversion{bold}$\ks\to3\po\kern8mm-$} \kern8mm
The decay $\ks\to3\po$ is purely CP violating. If CPT invariance holds, the BR for this decay is predicted to be BR\ab\pt1.9,-9,.
In KLOE, the signature is an event with 
a \kl\ crash, six photon clusters, and no tracks from the interaction
point. Background is mainly from $\ks\to\pio$ events with 
two spurious clusters from splittings or accidental activity.
Signal-event candidates are counted using the distribution in the plane
of two $\chi^2$-like discriminating variables, $\zeta_3$ and $\zeta_2$.
$\zeta_3$ is the quadratic sum of the residuals between the nominal
$\pi^0$ mass, $m_{\po}$, and the
invariant masses of three photon pairs formed from the six clusters 
present.
$\zeta_2$ is based on energy and momentum conservation in the
$\phi\to\ks\kl$, $\ks\to\pio$ decay hypothesis, 
as well as on the invariant masses of two photon pairs.
$\zeta_3$ and $\zeta_2$ are evaluated with the most favorable cluster
pairing in each case.

From 450~pb$^{-1}$ of data, and using the predicted distribution from an MC sample with an effective statistics of 5.3 times that of the data, \klo\ obtained $\zeta_2$ versus $\zeta_3$ distributions. The number of predicted background counts is $3.1\pm0.8\pm0.4$; while two counts in the signal box are observed in data.
KLOE thus obtains the 90\% C.L. limit BR$\leq$\pt1.2,-7, \cite{ks3pi0}, at present the most stringent limit on this BR.


\FIG\depmiss   \TAB\klbr
\subsection{\kl\ decays}
The KLOE measurement of the semileptonic \kl\ BRs 
\cite{klbr} uses tagging to obtain absolute BRs: the presence
of a \kl\ is tagged by observation of a $\ks\to\pic$ decay. 
An important point is that four modes, $\pi^\pm e^\mp\nu$ ($K_{e3}$), 
$\pi^\pm\mu^\mp\nu$ ($K_{\mu3}$), \pic\po, and 3\po, account for more 
than 99.5\% of all decays. For the first three modes, two
tracks are observed in the DC, whereas for the $3\po$ mode, only photons 
appear in the final state. The analysis of two-track and all-neutral-particle
events is therefore different.

Two-track events are assigned to the three channels of interest by use of a single variable:
the smaller absolute value of the two possible values of 
$\Delta_{\mu\pi}=|{\bf p}_{\,\rm
miss}|-E_{\,\rm miss}$, where
${\bf p}_{\,\rm miss}$ and $E_{\,\rm miss}$ are the missing momentum and
energy in the \kl\ decay, respectively, and are evaluated 
assuming the decay particles are a
pion and a muon. 

Figure \hyref\depmiss, left, shows an example of a $\Delta_{\mu\pi}$
distribution. A total of approximately 13 million tagged \kl\ 
decays (328 pb\up{-1}) are used for the measurement of the BRs. 

The numbers of $K_{e3}$, $K_{\mu3}$, and $\pic\po$ decays are obtained 
separately for each of 14 run periods by fitting
the $\Delta_{\mu\pi}$ distribution with the corresponding MC-predicted
shapes.
The signal extraction procedure is tested using
particle-identification variables from the calorimeter. 

Figure \hyref\depmiss, right, shows the $\Delta_{e\pi}$ spectrum for events with 
identified electrons, together with the results of a fit using MC shapes.
The $K_{e3(\gamma)}$ radiative tail is clearly evident.
The inclusion of radiative processes in the simulation
is necessary to obtain an acceptable fit, as well
as to properly estimate the fully inclusive radiative rates.

\vbox{\hypertarget{\depmiss}\break\cl{\figboxc fig05;11.6;}
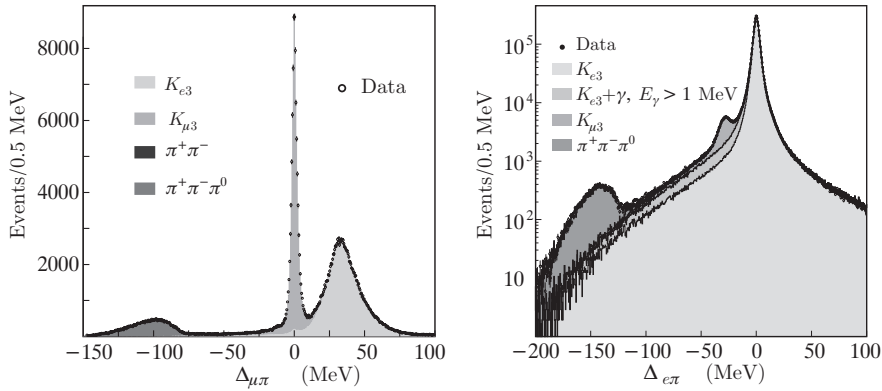
\allcap\depmiss;Left: distribution of $\Delta_{\mu\pi}$ for a run set. Right: distribution of $\Delta_{e\pi}$ for events with an 
identified electron, for the entire data set.;}
\noindent
The decay $\kl\to3\po$ is easier to identify. Detection
of $\ge3$ photons originating at the same point is accomplished with
very high efficiency (99\%) and very little background (1.1\%).

The errors on the absolute BRs are dominated by the 
uncertainty on the value of $\tau_{\kl}$, which enters into the 
calculation of the geometrical efficiency. This source of uncertainty
can be all but removed (at the cost of correlating the errors among
the BR measurements) by applying the 
constraint that the \kl\ BRs must add to unity.
The sum of the four BRs, plus the sum of the Particle Data Group
(PDG) values \cite{pdg06} 
for \kl\ decays to \pic, \pio, and $\gam\gam$ ($\sum$=0.0036), 
is $1.0104\pm0.0018\pm0.0074$.

Applying the constraint gives 
the results in Table \klbr.\vglue2mm
\def\eiii{\ifm{\pi^\pm e^\mp\nu}}  
\def\keiii{\ifm{K_{e3}}}
\def\muiii{\ifm{\pi^\pm \mu^\mp\nu}}   
\def\kmuiii{\ifm{K_{\mu3}}}
\begin{center}
\cl{Table \klbr. KLOE measurements of \kl\ branching ratios}\vglue2mm
\begin{tabular}{@{}lcccc@{}}
\hline\hline
\vst Mode & BR & $\delta$ stat & $\delta$ syst-stat & $\delta$ syst \\ 
\hline
\vst\eiii     &  0.4007  &  0.0005  &  0.0004  & 0.0014 \\
\vst\muiii    &  0.2698  &  0.0005  &  0.0004  & 0.0014 \\
\vst\pio\po   &  0.1997  &  0.0003  &  0.0004  & 0.0019 \\
\vst\pic\po   &  0.1263  &  0.0004  &  0.0003  & 0.0011 \\
\hline
\label{tab:finalbr}
\end{tabular}
\end{center}\vglue2mm\noindent
Constraining the BRs to add up to unity and solving for the
geometrical efficiency is equivalent to determining $\tau_{\kl}$ by
the second method described in section \ref{sec:KLlife}. KLOE finds
$\tau_{\kl} = 50.72\pm0.17\pm0.33$~ns.

For the ratio $\Gamma(K_{\mu3})/\Gamma(K_{e3})$,
KLOE obtains $R_{\mu e}=0.6734\pm0.0059.$
A value for comparison can be computed from the
slope of the $f_0$ form factor. We have also measured the vector form factor parameter \cite{lamlam}.



\FIG\kcmu
\subsection{\kpm\ decays}
\label{sec:kchdec}
The charged kaons, \kpm, decay mostly into $\mu^\pm\nu$ ($K_{\mu2}$) and $\pi^\pm\po$ ($K_{\pi2}$). The semileptonic modes account for about 8\% of all decays. Most dominantly, the $K_{\mu2}$ mode is \ab two-thirds of all \kpm\ decays. It is often used as a reference for measuring other BRs, and, together with the \kpm\ lifetime, provides a measurement of the decay constant $f_K$.

\vglue3mm\noindent{\mathversion{bold}$K^+\rightarrow\mu^+\nu(\gamma)\kern8mm-$} \kern8mm
The KLOE measurement of this BR \cite{kmu2}
is based on the use of $\km\to\mu^-\bar{\nu}$ decays for event tagging.
Identification of a $\km\to\mu^-\bar{\nu}$ decay requires the presence of a two-track vertex in the DC. The large number of $K_{\mu 2}$ decays allows for a statistical precision of \ab0.1\%, while setting aside a generous sample for systematic studies.
The shape of the $p^*$ distribution for signal events is obtained
from a sample of control data.
This distribution is used with the distributions for background 
sources to fit the $p^*$ spectrum ( Fig. \hyref\kcmu;, center panel).
Figure \hyref\kcmu; (right panel) shows the spectrum after background 
subtraction.
In a sample of four million tagged events,
KLOE finds \ab865,000 signal events with $225 \le p^* \le 400$ MeV, giving
${\rm BR}(\kp\to\mu^+\nu(\gam)) = 0.6366\pm0.0009\pm0.0015.$
This measurement is fully inclusive of 
final-state radiation (FSR) and has a 0.27\% uncertainty.

\vbox{\hypertarget{\kcmu}\break\cl{\figboxc fig09;12.4;}
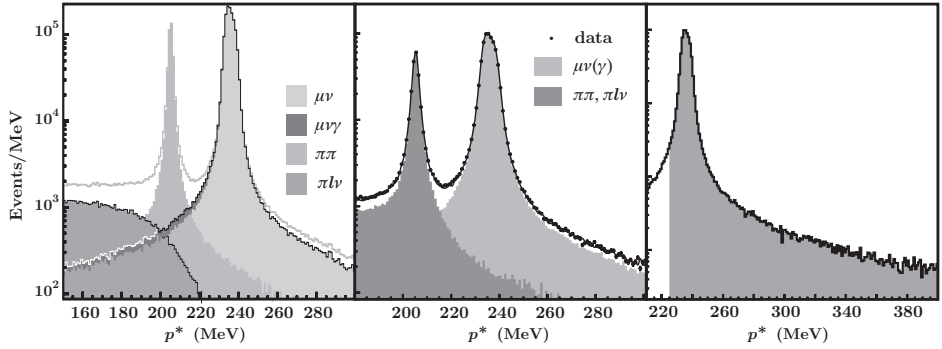
\allcap\kcmu;{({\it Left panel}\/) Monte Carlo spectra of $p^*$ for \kp\ decays, showing contributions from various channels. ({\it Center panel}\/) Distribution of $p^*$. ({\it Right panel}\/) Distribution of $p^*$ after background subtraction. The shaded area is used to count $\kp\to\mu^+\nu$ events.};}

\FIG\kcsl
\subsection{Semileptonic Decays of Charged Kaons}
The measurement is based on counting decays to each channel in samples
tagged by detection of two-body decays of the other kaon.
KLOE measures the semileptonic BRs separately for \kp\ and \km. Therefore, 
BR($\kpm\to\po e^\pm\nu(\gam)$) and BR($\kpm\to\po\mu^\pm\nu(\gam)$) are each determined from four independent measurements (\kp\ and \km\ decays tagged by $K\to\mu\nu$ and $K\to\pi\po$).
After removal of $\pi^+\pi^0$  decays from the signal sample, the leptonic \po\ is reconstructed from the two \gam s, providing the kaon decay point. Finally, the $m^2_\ell$ distribution for the decay lepton is reconstructed by TOF.
The $m^2_\ell$ distribution is shown in figure \hyref\kcsl;.\\
\vbox{\hypertarget{\kcsl}\break\cl{\figboxc fig10;6.3;}
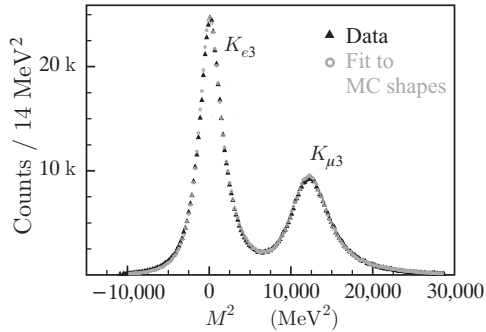
\allcap\kcsl;{Distribution of $m^2_\ell$, for events selected as $K^\pm_{\ell3}$ decays. Also shown is a fit to MC distributions for signal and background.};}\vglue1mm
\noindent
The four separate determinations of each BR provide
an estimate of the systematic uncertainties.
The KLOE values for the BRs and their ratio $R_{\mu e}$ are
$$\eqalign{
{\rm BR}(K^\pm_{e3})&=(5.047\pm0.046_{\rm stat+tag}\pm\sig_{\rm syst})\%\cr
{\rm BR}(K^\pm_{\mu3})&= (3.310\pm0.040_{\rm stat+tag}\pm\sig_{\rm syst})\%\cr
R_{\mu e}&= 0.656\pm0.008.\cr}$$
These results are preliminary, awaiting final determination of certain systematic uncertainties, still in progress \cite{eps}.

\section{$|V_{us}|$}
\label{kloevus}

In the Standard Model, the quark weak charged current is
\begin{displaymath}
J_\alpha^+=(\bar u\ \bar c\ \bar t)\gamma_\alpha(1-\gamma_5)\,{\bf V}
\pmatrix{d\cr s\cr b\cr},
\end{displaymath} 
where {\bf V} is a $3\times3$ unitary matrix introduced by
Kobayashi and Maskawa \cite{KM73} expanding on the original
suggestion by Cabibbo \cite{Cab63}.
The unitarity condition (${\bf V}^\dagger{\bf V}=1$) is required
by the assumption of universality of the weak interactions of leptons
and quarks and the absence of flavor-changing neutral currents.
The realization that a precise test of CKM unitarity can be obtained 
from the first-row constraint
$|V_{ud}|^2 + |V_{us}|^2 + |V_{ub}|^2 = 1$ (with $|V_{ub}|^2$ negligible)
has sparked a new interest in good measurements of quantities related to
$|V_{us}|$.
As we discuss in the following $|V_{us}|$ can be determined 
using semileptonic kaon decays;
the experimental inputs are the BRs, lifetimes,
and form-factor slopes. Both neutral (\ks\ or \kl) and charged kaons
may be used and provide independent measurements.

KLOE is unique in that it is the only experiment that
can by itself measure the complete set of experimental inputs for the
calculation of $|V_{us}|$ using both charged and neutral kaons.
Recall that we are at a \f\ factory thus are especially suited for measurements of the \kl\ and \kpm\ lifetimes.
In addition, KLOE is the only experiment that can measure \ks\
BRs at the sub-percent level.

\subsection{Semileptonic Kaon Decays}
\label{sec:vusl3}
The semileptonic kaon decay rates
still provide the best means for the measurement of $|V_{us}|$
because only the vector part of the weak current contributes
to the matrix element $\bra\pi;J_\alpha\ket K;$. For vector
transitions, the Ademollo-Gatto theorem \cite{ag} ensures that
$SU(3)$ breaking appears only to second order in $m_s-m_{u,d}$. Still one has to properly account for radiative correction, $SU(2)$ and $SU(3)$ corrections, \cite{marros}. Finally, the matrix element of the weak vector current contains two form factor, FF, which reflect on the value of the phase space integral. For this last point we have no theoretical help but we can measure their shape experimentally.
The semileptonic decay rates, fully inclusive of radiation, are given by
$$
\Gamma_i(K_{e3,\,\mu3})=|V_{us}|^2\:{C_i^2\:G^2\:M^5\over768\pi^3}\:S_{\rm
EW}\:(1+\delta_{i,\,\rm em}+
\delta_{i,\,SU(2)})\:|f^{\ko}_+(0)|^2\:I_{e3,\,\mu3}.\eqn\gamsl$$
In the above expression, $i$ indexes \ko\ and $K^\pm$. 
$S_{\rm EW}$ is the universal short-distance radiative correction factor and the $\delta$ terms are all the other correction mentioned.
The phase space integrals  $I_{e3,\,\mu3}$ require knowledge of the FF parameters. The numerical factor
in the denominator of eq. \hyrefq\gamsl, 768=3\x2\up8, is chosen in
such a way that $I=1$ when the masses of all final-state particles vanish.
For $K_{e3}$, $I \approx 0.56$ and for $K_{\mu3}$, $I \approx 0.36$.

The biggest uncertainty at present is in the value of the vector form factor at zero momentum transfer, $f_+^{\ko}(0)$ in eq. \hyrefq\gamsl. Progress is expected from lattice calculations. It has therefore become customary to extract from data and eq. \hyrefq\gamsl\ the value of $\kappa=f_+^{\ko}(0)\,|V_{us}|$, which contains all the measured parameters described above. 

KLOE finds $\kappa=0.2157\pm0.0006,\ 0.2163\pm0.0008,\ 0.2155\pm0.0014,\ 0.2171\break\pm0.0021, 0.2150\pm0.0028$, respectively, for \kl($e$3), \kl($\mu$3), \ks($e$3), \kpm($\mu$3), \kpm($e$3) using exclusively KLOE data except for the slope of the scalar FF, $\lambda_0$ \cite{ktis}, which we are currently measuring and the \ks\ lifetime which is very well known since a long time \cite{pdg06}. The above value are quite consistent with each other. The weighted average is $\langle\kappa\rangle=0.2160\pm0.005$. This value is in agreement with more limited results from other experiments. It compares well with the value $\sqrt{1-|V_{ud}|^2}\x f_+^{\ko}(0)$=0.2187\plm0.0022, using $f_+^{\ko}(0)$=0.961\plm0.008 (Leutwyler and Ross \cite{marros}) and $|V_{ud}|$=0.9740\plm0.0003 (Marciano and Sirlin, \cite{marros}). In conclusion,Unitarity of the CKM mixing matrix, requiring $|V_{ud}|^2+|V_{us}|^2$=1.0 to better than 0.00002, given the smallness of $|V_{ub}|$, is essentially satisfied.

\FIG\allckm
\subsection{$K\to\mu\nu$ decays}
Reliable lattice quantum chromodynamics (QCD) results have 
recently become available and are rapidly improving \cite{lat}. The availability of precise values for the pion- and kaon-decay constants $f_\pi$ and $f_K$ allows the use of a relation between $\Gamma(K_{\mu2})/\Gamma(\pi_{\mu2})$ and $|V_{us}|^2/|V_{ud}|^2$, with the advantage that lattice-scale uncertainties and radiative corrections largely cancel out in the ratio \cite{marfk}:
$${\Gamma(K_{\mu2(\gamma)})\over\Gamma(\pi_{\mu2(\gamma)})}=%
{|V_{us}|^2\over|V_{ud}|^2}\;{f_K^2\over f_\pi^2}\;%
{m_K\left(1-m^2_\mu/m^2_K\right)^2\over m_\pi\left(1-m^2_\mu/m^2_\pi\right)^2}\x(0.9930\pm0.0035).\eqn\fkfp$$
All information about the unitarity of the CKM matrix, as verified on the first row, is represented graphically in fig. \hyref\allckm.\vglue1mm

\vbox{\hypertarget{\allckm}\break\cl{\figboxc vus-vud;7;}
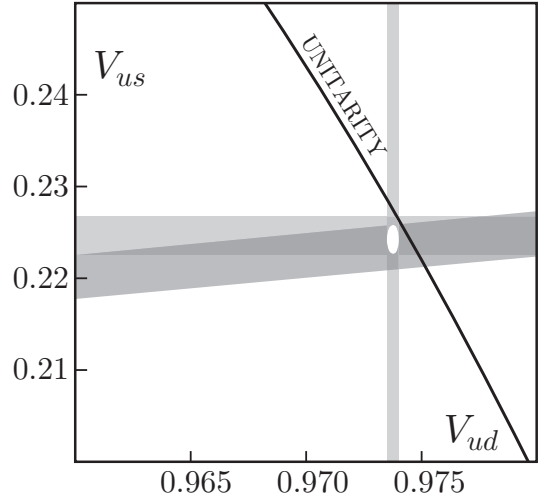
\allcap\allckm;The vertical band gives the value of $|V_{ud}|$ as measured in nuclear $\beta$-decay. The horizontal band gives $|V_{us}|$ from semileptonic kaon decays, using $f_+^{\ko}(0)$=0.96\plm 0.008. The slightly inclined band follows from eq. \hyrefq\fkfp\ and the KLOE result for BR($K_{\mu2}$). Unitarity is represented by an arc of circle. The white ellipses is the 1 \sig\ contour from a fit to all data;}

\FIG\interf
\section{Quantum Interferometry}
\label{sec:inter}
As stated in section \ref{ksbeam}, the neutral kaon pair $\f\to\ko\kob$
are produced in a pure quantum state with $J^{PC}$=1\up{--}. We evolve the initial state in  eq. \hyrefq\twok\ in time, project to any two 
possible final states $f_1$ and $f_2$, take the modulus
squared, and integrate over all $t_1$ and $t_2$ for fixed $\Delta t=t_1-t_2$
to obtain (for $\Delta t > 0$, with $\Gamma\equiv\Gamma_L+\Gamma_S$)
$$
\eqalign{
I_{f_1,\,f_2}(\Delta t)&=
{1\over2\Gamma}|\bra f_1;\ks\rangle\bra f_2;\ks\rangle|^2 \x
\Big[|\eta_1|^2e^{-\Gamma_L\Delta t}+|\eta_2|^2e^{-\Gamma_S\Delta t}\cr
& -2(1-\zeta)|\eta_1||\eta_2|
e^{-(\Gamma_L+\Gamma_S)\Delta t/2}\cos(\Delta m\Delta t+\f_2-\f_1)\Big].\cr}\eqn\eqtwo$$
The last term is due to interference between the decays to states $f_1$ and $f_2$ and $\zeta$=0 in quantum mechanics. Fits to the $\Delta t$ distribution provide measurements of the magnitudes and phases of the parameters $\eta_i = \bra f_i;\kl\rangle/\break\bra f_i;\ks\rangle$, as well as of the \kl-\ks\ mass difference $\Delta m$ and the decay rates $\Gamma_L$ and $\Gamma_S$.

Such fits also allow tests of fundamental properties of quantum mechanics. For example, the persistence of quantum-mechanical coherence can be tested by choosing $f_1 = f_2$. In this case, because of the antisymmetry of the initial state and the symmetry of the final state, there should be no events with $\Delta t = 0$. 
Using the 2001--2002 data, KLOE has conducted a preliminary analysis of 
the $\Delta t$ distribution for $\ks\kl\to\pic\pic$ events
that establishes the feasibility of such tests \cite{interf}.
The $\Delta t$ distribution is fit with a function of the form of eq. \hyrefq\eqtwo, including the experimental resolution and a peak from
$\kl\to\ks$ regeneration in the beam pipe. The results are shown in
figure \hyref\interf.
Observation of $\zeta\neq0$ would imply loss of quantum coherence. The value of $\zeta$ depends on the basis, \ko-\kob\ or \ks-\kl\ used in the analysis. Using the \ko-\kob\ basis we find $\zeta$=\pt(0.1\pm0.2\pm0.04),-5, \ie\kern3mm no violation of quantum mechanics.\vglue1mm
\vbox{\hypertarget{\interf}\break\cl{\figboxc fig04;8.3;}\vglue.1mm
\allcap\interf;$\Delta t$ distribution for $\f\to\ks\kl\to\pic\pic$
events.;}
\FIG\delm
\section{Test of CPT}
The three discrete symmetries of quantum mechanics, charge conjugation ($C$), parity ($P$) and time reversal ($T$) are known to be violated in nature, both singly and in pairs. Only $CPT$ appears to be an exact symmetry of nature. Exact $CPT$ invariance holds in quantum field theory which assumes Lorentz invariance (flat space), locality and unitarity~\cite{CPT}. These assumptions could be violated at very large energy scales, where quantum gravity cannot be ignored~\cite{CPTV}. Testing the validity of $CPT$ invariance probes the most fundamental assumptions of our present understanding of particles and their interactions.
The neutral kaon system offers unique possibilities for the study of $CPT$ invariance. From the requirement of unitarity, Bell and Steinberger have derived a relation, the so called Bell-Steinberger relation (BSR)~\cite{BS:paper}. The BSR relates a possible violation of $CPT$ invariance ($m_{\ko}\neq m_{\kob}$ and/or $\Gamma_{\ko}\neq\Gamma_{\kob}$) in the time-evolution of the $\ko-\kob$ system to the observable $CP$-violating interference of \ks\ and \kl\ decays into the same final state $f$, in terms of $CP$- and $CPT$-violating parameters  $\Re\eps$ 
and $\Im\delta$. Recently we succeeded in improving the determination of those two parameters using \klo\  results. 
Specifically we benefitted from our new precise measurement of the branching ratio for \kl\ decays to \pic \cite{klpipi}, which is relevant for the determination of  $\Re\eps$. We improved the upper limit on \BR(\ks\to\pio\po) \cite{ks3pi0}, which is necessary to improve the accuracy on  $\Im\delta$. Finally we obtained the first measurement of the charge asymmetry $A_S$ for \ks\ semileptonic decays. The latter allowed us to determine the direct contribution from semileptonic channels {\it without assuming unitarity}. Our study resulted in narrowing the mass difference to \pt(-5.3\leq m_{\ko}-m_{\kob}\leq6.3),-19, GeV \@\ 95\% CL, see fig. \hyref\delm, a factor of two better than previous limits \cite{BSRpaper}.

\vbox{\hypertarget{\delm}\break\cl{\figboxc fig2a;12;}
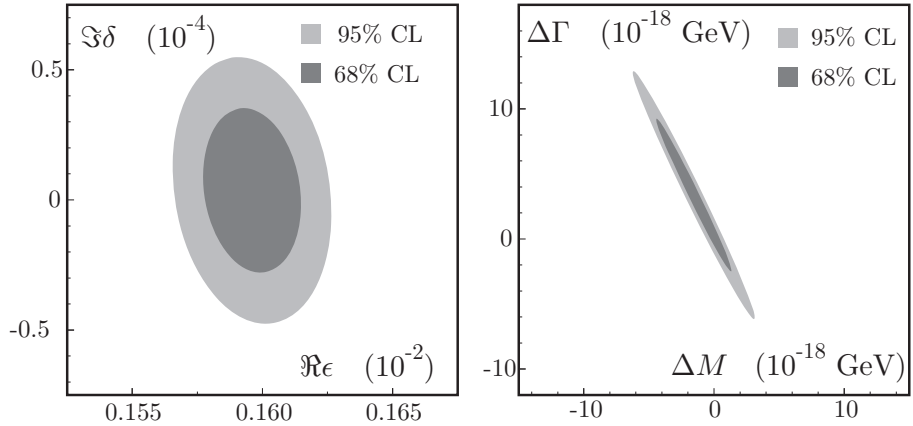
\allcap\delm;Left: allowed region at 68\% and 95\% C.L. in the $\Re\eps$, $\Im\delta$ plane. Right: allowed region at 68\% and 95\% C.L. in the $\Delta M, \Delta \Gamma$ plane.;}\vglue1mm

\section {\f\ radiative decays}

It is serendipitous that while on the \f\ peak kaons are produced furiously, it   also happens that, true its its name, \DAF\ is a \ff\ . These \f\ mesons, by electromagnetic transitions ($\f\to\mbox{meson}+\gam$) to other mesons, result in a great profusion of lighter scalar and pseudoscalar mesons. The transition rates are strongly dependent on the wave function of the final-state meson. They also depend on its flavor content, because the \f\ is a nearly pure $s\bar{s}$ state and because there is no photon-gluon coupling. These radiative \f\ decays are unique probes of meson properties and structure. This hadron source, coupled with \klo\ 's great versatility, allowed \klo\ to perform many hadronic experiments, with far greater precision than has been done elsewhere before.

In addition, initial-state radiation (ISR) lowers the effective collision energy squared from $s$ to ${s_\pi} = (s - 2E_\gamma\sqrt{s})$, providing access to hadronic states of mass $\sqrt{s_\pi}$ from threshold to $m_\f$.
KLOE has exploited ISR to measure the $\epem\to\pic$ cross section over this entire energy range without changing the \DAF\ energy \cite{sighad}. 
In this paper we have no space to delve into this last fascinating and somewhat fractious topic. Similarly, due to space constraint, we will limit ourselves to mentioning only a handful of the radiative decays measurements, just enough to give a taste of hadron physics with \klo.\vglue3mm   

\noindent{\bf\mathversion{bold}The $\eta$ Meson$\kern8mm-$} \kern8mm
The BR for the decay $\f\to\eta\gamma$
is 1.3\%. In 2.5~fb$^{-1}$ of KLOE data, there are 100 million $\eta$ 
mesons, identified clearly by their recoil against a photon of $E = 363$~MeV\@.
\klo\ has used the $\eta$ meson to test discrete symmetry laws in several ways.
1.- The symmetries of the Dalitz plot for $\eta\to\pic\po$ decays
provide tests of charge-conjugation invariance. \klo\ has made such plots with about 1.4 million events and obtain violation limits of the order of $0.002\pm 0003$.
2.- The decay $\eta\to3\gam$ violates \C, while the decay $\eta\to\pic$ violates both \P\ and \CP. KLOE has conducted a search for the decay $\eta\to3\gam$ using 410~pb$^{-1}$ and obtained the limit ${\rm BR}(\eta\to3\gam)\leq\pt1.6,-5,$ at 90\% C.L., the most stringent obtained to date \cite{eta3g}.
3.- KLOE has also searched for evidence of the decay $\eta\to\pic$ 
in the tail of the $M_{\pi\pi}$ distribution for
$e^+e^-\to\pic\gam$ events in which the photon is emitted at
large polar angles ($\theta>45\deg$).
No peak is observed in the distribution of $M_{\pi\pi}$ in
the vicinity of $m_\eta$. The corresponding limit is ${\rm BR}(\eta\to\pic)\leq\pt1.3,-5,$ at 90\% C.L. \cite{etapp}, which is more stringent than the previous limit by a factor of 25.

The decay $\eta\to\po\gam\gam$ is particularly interesting in chiral
perturbation theory. There is no $\mathcal{O}(p^2)$ contribution, and the $\mathcal{O}(p^4)$ contribution is small. At KLOE, the decay can be reconstructed with full kinematic closure and without complications from certain backgrounds present in fixed-target experiments, such as $\pi^-p\to\pio n$. 
Using one fifth of its data, \klo\ has obtained the preliminary 
result ${\rm BR}(\eta\to\pi^0\gam\gam) = \pt(8.4\pm2.7\pm1.4),-5,$,
which is in agreement with several $\mathcal{O}(p^6)$
chiral perturbation theory calculations discussed in a recent $\eta$ workshop \cite{wasa}. The final measurement using the complete set of data is expected in 2007.\vglue3mm   

\noindent{\bf\mathversion{bold}The $\eta'$ Meson$\kern8mm-$} \kern8mm
The magnitude of ${\rm BR}(\f\to\eta'\gam)$
is a probe of the $s\bar{s}$ content of the $\eta'$. Furthermore 
the ratio $R_{\phi} \equiv {\rm BR}(\f\to\eta'\gam)/{\rm BR}(\f\to\eta\gam)$
can be related to the pseudoscalar mixing angle
$\varphi_{\rm P}$ in the basis 
$\{\ket u\bar u+d\bar d;/\sqrt2,\ \ket s\bar s;\}$,
offering an important point of comparison
for the description of the $\eta$-$\eta'$ mixing in extended 
chiral perturbation theory \cite{Fel00}.

At \klo\ the $\f\to\eta'\gam$ decay is identified in either the charged pion channel where $\eta'\to\eta\pic$ with $\eta\to\gam\gam$ or with $\eta\to 3\po$, or in the neutral pion channel $\eta'\to\eta\pio$ with $\eta\to\pic\po$.
The $\f\to\eta\gam$ decay is identified in the channel in which 
$\eta\to\pic\po$.
To date, from study of one fifth of the \klo\ data collected, we obtained $R_{\phi}=(4.77\pm0.09_{stat}\pm0.19_{sys})\x 10^{-3}$ from which we derive 
$BR(\phi \to \eta^{\prime} \gamma)=(6.20\pm0.11_{stat}\pm0.25_{sys})\x 10^{-5}$.
Scant knowledge of intermediate branching ratios prevents us to benefit from having twentyfold statistics over our previous BR determination \cite{etapr}.
In the hypothesis of no gluonium content we extract the pseudo-scalar mixing angle in the quark-flavor basis $\varphi_P=(41.4\pm0.3_{stat}\pm0.7_{sys}\pm0.6_{th})^{\circ}$. Combining our $R_{\phi}$ result with other experimental constraints, we estimate
the gluonium fractional content of $\eta^{\prime}$ meson as $Z^2 = 0.14\pm0.04$ and the mixing angle $\varphi_P = (39.7\pm0.7)^{\circ}$ \cite{didonato}.

\FIG\kloop    \FIG\fopp
\subsection{Scalar Mesons: $f_0(980)$ and $a_0(980)$}
\label{sec:f0}
The compositions of the five scalar mesons with masses below 1~GeV are not well understood. In fact, a detailed proposal of how to elucidate their nature using \klo\ at \DAF\ was one of the first papers JLF had written, after her arrival to LNF in 1991 \cite{JLF}. Theoretical models have proliferated greatly, we only pick two which are illustrated below. \klo\  has now collected literally millions of such decays. We not only measure branching ratios, we can obtain a very densily populated Dalitz plot. The two dimensional Dalitz plot density, subdivided into fine cells, can be fit to theoretical models. As of this writing, some theoretical models are so over encumbered with parameters that the physical meaning of several are opaque, and render fits to them difficult to interpret. So here we leave out all detailed discussions of the recent \klo\ fits. 
 
{\bf Branching Ratios}: \ the branching ratio for the decays  $\f\to f_0\gam\to\pi\pi\gam$ and $\f\to a_0\gam\to\eta\po\gam$ are suppressed unless the $f_0$ and $a_0$ have significant $s\bar{s}$ content. The BRs for these decays are estimated to be on the order of $10^{-4}$ if the $f_0$ and $a_0$ are $qq\bar{q}\bar{q}$ states (in which case they contain an $s\bar{s}$ pair), or on the order of $10^{-5}$ if the $f_0$ and $a_0$ are conventional $q\bar{q}$ states. If the $f_0$ and $a_0$ are $K\Kb$ molecules, the BR estimates are sensitive to assumptions about the spatial extension of these states.
 
{\bf Invariant Mass Distributions}: \ $\pi\pi$
and $\eta\pi^0$ invariant-mass distributions can shed light on the nature of the $f_0$ and $a_0$ because fits to the distributions provide estimates of the couplings of these mesons to the \f\ and/or to the final-state particles.
To make fits one necessarily assume a specific model for the decay mechanism. Because of the proximity of the $f_0$ and $a_0$ masses to the $K\Kb$ threshold, and because these mesons are known to couple strongly to $K\Kb$, the kaon-loop model, figure \hyref\kloop, is often used \cite{AI89}. 

\vbox{\hypertarget{\kloop}\break\cl{\figboxc fig12;8.9;}
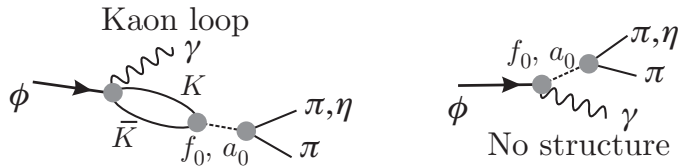
\allcap\kloop;{Amplitudes used to compute the $\pi\pi$ and $\eta\po$ mass spectra
for $\f\to S\gam$.};} 
\noindent
The expression for the rate for E1 $\f\to S\gam$ transitions 
($S = f_0$ or $a_0$) contains a factor $E_\gam^3$ (where $E_\gam$
is the energy of the radiated photon), as required by considerations of
phase space and gauge invariance.
As a result, the invariant-mass distributions in $\f\to S\gam$ 
decays are cut off above
$m_\f$ and develop a long tail toward lower mass values (see figure \fopp, right).
The fit results therefore depend strongly on the scalar meson masses and widths.
\vglue3mm   

\noindent{\bf\mathversion{bold}$\f\to\pi^0\pi^0\gam$ and $\f\to\eta\pi^0\gam\kern8mm-$} \kern8mm
The first KLOE studies of the decays $\f\to\pio\gam$
\cite{f0} and $\f\to \eta\po\gam$ \cite{a0} were performed with the 
17~pb$^{-1}$ of data collected in 2000. All Br's quoted therein are in agreement with, but are much more precise than previous measurements.

The amplitudes contributing to $\f\to\pio\gam$ include $\f\to S\gam$
(with $S = f_0$ or $\sigma$) and $\f\to\rho^0\po$ (with $\rho^0\to\po\gam$).
The $M_{\pi\pi}$ distribution was
fit with a function including terms describing the contribution from $\f\to
S\gam$, that from $\f\to\rho\pi$, and their interference. 
The value obtained for ${\rm BR}(\f\to\pio\gam)$ was 
\pt(1.09\pm0.03\pm0.05),-4,. In the $a_0$ case, the amplitudes contributing are $\f\to a_0\gam$ and $\f\to\rho^0\po$, with $\rho^0\to\eta\gam$.
In KLOE analyses, final states corresponding to
two different $\eta$ decay channels were studied:
$\eta\to\gam\gam$ ($5\gam$ final state) and
$\eta\to\pic\po$ ($\pic5\gam$ final state).
The $M_{\eta\pi}$ distributions for each sample
were fit simultaneously with a function analogous to that discussed
above. The values obtained for ${\rm BR}(\f\to\eta\po\gam)$
using the $5\gam$ and $\pic5\gam$ samples were 
\pt(8.51\pm0.51\pm0.57),-5, and \pt(7.96\pm0.60\pm0.40),-5,, respectively.

The analyses of the $\f\to\pio\gam$ and $\f\to\eta\po\gam$ decays
based on a later data set, benefit from an increase in statistics
by a factor of \ab30 such that one could eschew background removal using severe cuts which could introduce biases, and plot all events with the same topology in Dalitz plots. In such cases one fits the whole Dalitz plots by correctly modelling all channels in the MC, fold in cell by cell efficiencies, then extract the relative fraction of each channel and relevant parameters, by fitting to theoretical models. In particular, we fit to both 1, the much, much improved k-loop model \cite{AK05}, and also 2, to the no-structure model \cite{IMP06} illustrated in figure \kloop; (right panel). At present since the interpretations of some parameters, as well as the stability of some theoretical fits, are under dicussion between \klo\ and the theorists, we only quote here the BR to $\f\to\pio\gam$ \cite{simona}:
$${\rm BR}(\f\to S\gam \to \pio\gam)=\pt\left(1.07^{+0.01}_{-0.04}{\rm(fit)}\ ^{+0.04}_{-0.02}{\rm(syst)}\ ^{+0.06}_{-0.05} {\rm (mod)}\right),-4,.$$
\vglue3mm   

\noindent{\bf\mathversion{bold}$\f\to\pi^+\pi^-\gam\kern8mm-$} \kern8mm
KLOE has also published a study of the decay $\f\to f_0\gam\to\pic\gam$
\cite{f0ppg}. The $f_0$'s charged mode was extremely difficult to identify because only a small fraction of the $e^+e^-\to\pic\gam$ events involve the $f_0$; the principal contributions are from events in which the photon
is from ISR or FSR. The analysis is performed on the $M_{\pi\pi}$ distribution for events with large photon polar angle where the ISR contribution is much reduced. The $M_{\pi\pi}$ distribution was fit with a function composed of analytic expressions describing  the ISR, FSR, $\rho\pi$ contributions and two terms that describe the decay $\f\to S\gam\to\pi\pi\gam$ and its 
interference with FSR which could be constructive or destructive.

Figure \fopp, left, shows the $M_{\pi\pi}$ distribution, 
with the result of the kaon-loop fit superimposed. The overall appearance of the distribution is dominated by the radiative return to the $\rho$; the $f_0$ appears as
the peak-like structure in the region 900--1000 MeV. 
Figure \hyref\fopp, right, shows the data in this
mass region, with the ISR, FSR, and $\rho\pi$ terms subtracted.\vglue1mm 
\vbox{\hypertarget{\fopp}\break\cl{\figboxc fig13;11.5;}
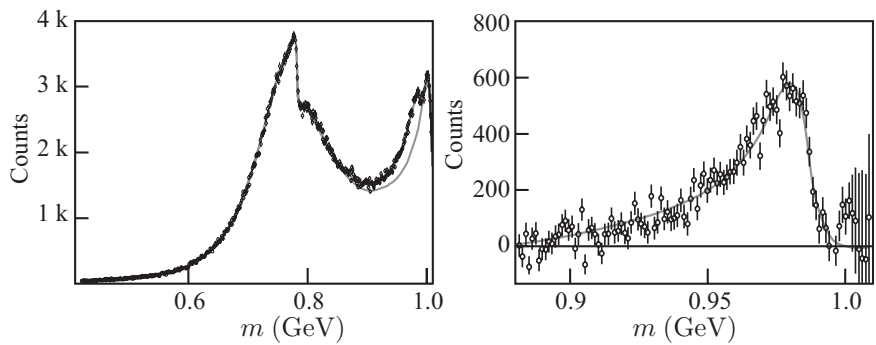
\allcap\fopp;{Left: $M_{\pi\pi}$ distribution for large-photon-angle $\pic\gam$ events from KLOE. Fits with and without the $f_0$ contribution are shown. Right:  Distribution in the region of the $f_0$, with the initial-state radiation, final-state radiation, and $\rho\pi$ contributions subtracted.};}\vglue1mm
\noindent
Both the kaon-loop and no-structure fits strongly prefer destructive interference between the $S\gam$ and FSR amplitudes. 
The kaon-loop fit gives coupling values in reasonable agreement
with those obtained from the fit to the KLOE $\f\to\pio\gam$ data
discussed above. Integrating the appropriate terms of the fit functions gives values for ${\rm BR}(\f\to f_0\gam)$ in the neighborhood of \pt3,-4,.

\vspace{3ex}
\cl{\bf\large Acknowledgements}\vglue3mm

We thank the organizers, Lia Panchieri and Maria Krawczyk, of the Final Euridice Meeting at Kazimierz, August 2006, for inviting us to contribute a pedagogical paper on KLOE in their proceedings.  This is a token of our appreciation of the almost decade-long symbiosis between Euridice, Eurodaphne, and KLOE. 

We hope that our fellow KLOE collaborators forgive us for our simplified discussion of their hard and elegant labors. Finally, one of us, JLF, wants especially to thank Caterina Bloise for representing KLOE at Lia Fest which celebrated so joyfully our friendship and collaborations.


\begin{thebibliography}{999}

\newcommand{\EPJA}{{Eur.\ Phys.\ J.\ A}}
\newcommand{\EPJC}{{Eur.\ Phys.\ J.\ C}}
\newcommand{\JHEP}{{JHEP}}
\newcommand{\JPG}{{J.\ Phys.\ G}}
\newcommand{\NIMA}{{Nucl.\ Instrum.\ Methods A}}
\newcommand{\NPA}{{Nucl.\ Phys.\ A}}
\newcommand{\NPB}{{Nucl.\ Phys.\ B}}
\newcommand{\PLB}{{Phys.\ Lett.\ B}}
\newcommand{\PRC}{{Phys.\ Rev.\ C}}
\newcommand{\PRD}{{Phys.\ Rev.\ D}}
\newcommand{\PRep}{{Phys.\ Rep.}}
\newcommand{\PRev}{{Phys.\ Rev.}}
\newcommand{\PRL}{{Phys.\ Rev.\ Lett.}}
\newcommand{\PTP}{{Prog.\ Theor.\ Phys.}}
\newcommand{\RMP}{{Rev.\ Mod.\ Phys.}}
\newcommand{\ZPC}{{Z.\ Phys.\ C}}

\bibitem{DAFNE+90:prop}
Preprint LNF-90/031(R), (1990). http://www.lnf.infn.it/sis/preprint.

\bibitem{DCNIM}
M.~Adinolfi {\it et al.}, (KLOE Collab.), Nucl.\ Instrum.\ Meth.\ A {\bf 488} (2002) 51.

\bibitem{EmCNIM}
M.~Adinolfi {\it et al.}, (KLOE Collab.), Nucl.\ Instrum.\ Meth.\ A {\bf 482} (2002) 364.

\bibitem{trigNIM}
M.~Adinolfi {\it et al.}  (KLOE Collab.),
Nucl.\ Instrum.\ Meth.\ A {\bf 492} (2002) 134.

\bibitem{lumi}
F.~Ambrosino {\it et al.}, (KLOE Collab.),
Eur.\ Phys.\ J.\ C {\bf 47} (2006) 589.

\bibitem{DAQNIM}
A.~Aloisio {\it et al.}, (KLOE Collab.),
Nucl.\ Instrum.\ Meth.\ A {\bf 516} (2004) 288.

\bibitem{offlineNIM}
F.~Ambrosino {\it et al.}, (KLOE Collab.),
Nucl.\ Inst.\ Meth.\ A {\bf 534} (2004) 403.

\bibitem{pdg06}
W.-M. Yao {\it et al.}, (Particle Data Group), J. Phys. G 33, 1 (2006).

\bibitem{llwidth}
F.~Ambrosino {\it et al.}, (KLOE Collab.),
Phys.\ Lett.\ B {\bf 608} (2005) 199.

\bibitem{kllife}
F.~Ambrosino {\it et al.}, (KLOE Collab.), Phys.\ Lett.\ B {\bf 626} (2005) 15.

\bibitem{klbr}
F.~Ambrosino {\it et al.}, (KLOE Collab.),
Phys.\ Lett.\ B {\bf 632} (2006) 43.

\bibitem{ks2pi}
A.~Aloisio {\it et al.}, (KLOE Collab.),
Phys.\ Lett.\ B {\bf 537} (2002) 21; 
F.~Ambrosino {\it et al.} (KLOE Collab.), \EPJC\ {\bf48} (2006)  767.

\bibitem{kslept}
F.~Ambrosino {\it et al.}  (KLOE Coll.), Phys.\ Lett.\ B {\bf 535} (2002) 37;
Phys.\ Lett.\ B {\bf 636} (2006) 173.

\bibitem{ks3pi0}
F.~Ambrosino {\it et al.}  (KLOE Collab.),
Phys.\ Lett.\ B {\bf 619} (2005) 61.

\bibitem{lamlam}
F.~Ambrosino {\it et al.}, (KLOE Collab.),
Phys.\ Lett.\ B {\bf 636} (2006) 166.

\bibitem{kmu2}
F.~Ambrosino {\it et al.}, (KLOE Collab.),
Phys.\ Lett.\ B {\bf 632} (2006) 76.

\bibitem{eps}
F.~Ambrosino {\it et al.}, (KLOE Collab.),
PoS {\bf HEP2005} (2006) 287.

\bibitem{KM73}
M. Kobayashi, T. Maskawa, \PTP\ {\bf652} (1973) 49.

\bibitem{Cab63}
N. Cabibbo, \PRL\ {\bf 531} (1963) 10.

\bibitem{ag}
M. Ademollo R. Gatto, \PRL\ {\bf264} (1964) 13.

\bibitem{marros}
A. Sirlin, \RMP\/ {\bf573} (1978) 50; \NPB\ {\bf83} (1982) 196. H. Leutwyler, M. Roos, \ZPC\ {\bf91} (1984) 25. V. Cirigliano et al. \EPJC\ {\bf21} (2002) 23. W.J. Marciano and A. Sirlin, \PRL\ {\bf 96} (2006) 032002.

\bibitem{ktis}
There is still some confusion about the value of $\lambda_0$. We use preliminary values from CKM06, Nagoya.

\bibitem{lat}
C.T.H. Davies, et al. \PRL\/ {\bf92} (2004) 022001.

\bibitem{marfk}
W.J. Marciano, \PRL\ {\bf93} (2004) 231803.

\bibitem{interf}
F.~Ambrosino {\it et al.}, (KLOE Collab.),
Phys.\ Lett.\ B {\bf 642} (2006) 315.

\bibitem{CPT}
  G.~Lueders,
   Annals Phys.\  {\bf 2} (1957) 1
  [Annals Phys.\  {\bf 281} (2000) 1004].

\bibitem{CPTV}
  See e.g.~V.A.~Kostelecky and R.~Lehnert,
  Phys.\ Rev.\ D {\bf 63} (2001) 065008;
  N.~E.~Mavromatos,
  AIP Conf.\ Proc.\  {\bf 796} (2005) 13.

\bibitem{BS:paper}J.S. Bell and J. Steinberger, {\it Proc. Oxford
Int. Conf. on Elementary Particles} (1965).

\bibitem{klpipi}
F.~Ambrosino {\it et al.}, (KLOE Collab.),
Phys.\ Lett.\ B {\bf 638} (2006) 140.

\bibitem{BSRpaper}
F. Ambrosino {\it et al.},  (KLOE Collab.), JHEP {\bf 0612} (2006) 011.

\bibitem{sighad}
A.~Aloisio {\it et al.}, (KLOE Collab.),
Phys.\ Lett.\ B {\bf 606} (2005) 12.

\bibitem{eta3g}
A.~Aloisio {\it et al.}, (KLOE Collab.),
Phys.\ Lett.\ B {\bf 591} (2004) 49.

\bibitem{etapp}
F.~Ambrosino {\it et al.}, (KLOE Collab.),
Phys.\ Lett.\ B {\bf 606} (2005) 276.

\bibitem{wasa}
Eta Physics Handbook, J. Bijnens J, G. Galt, B.M.K. Nefkens, eds., {\it Proc.\
Workshop Eta Phys., Uppsala, 2001.} Phys. Scripta {\bf T99} (2002).

\bibitem{Fel00}
Th. Feldmann, Int.\ J.\ Mod.\ Phys.\ A {\bf159} (2000) 15.

\bibitem{etapr}
A.~Aloisio {\it et al.}, (KLOE Collab.),
Phys.\ Lett.\ B {\bf 541} (2002) 45.

\bibitem{didonato}
Ambrosino F, {\it et al.} (KLOE Collab.) hep-ex/0612029, \PLB\ (2007) in print.

\bibitem{JLF}
P.~J.~Franzini, W.~Kim and J.~Lee-Franzini,
Phys.\ Lett.\ B {\bf 287} (1992) 259.

\bibitem{AI89}
N.N. Achasov, V.N. Ivanchenko \NPB\ {\bf315} (1989) 465.

\bibitem{f0}
A.~Aloisio {\it et al.}, (KLOE Collab.),
Phys.\ Lett.\ B {\bf 537} (2002) 21.

\bibitem{a0}
A.~Aloisio {\it et al.}, (KLOE Collab.),
Phys.\ Lett.\ B {\bf 536} (2002) 209.

\bibitem{AK05}
N.N. Achasov, A.V. Kiselev, \PRD\ {\bf73} (2006) 054029.

\bibitem{IMP06}
G. Isidori, {\it et al.}, \JHEP\ {\bf 0605} (2006) 049.

\bibitem{simona}
F. Ambrosino {\it et al.}, (KLOE Collab.) arXiv:hep-ex/0609009, EPJC {2006} in print.

\bibitem{f0ppg}
F.~Ambrosino {\it et al.}, (KLOE Collab.),
Phys.\ Lett.\ B {\bf 634} (2006) 148.



\end{thebibliography}
\end{document}